\documentstyle[12pt]{article}
\begin{document}

\begin{center}

{\Large \bf Separable Four-dimensional \\[0.3ex]
Harmonic Oscillators and \\[1ex]
Representations of the Poincar\'e Group} \\[4ex]

{\large \bf Y. S. Kim}\\
{\large  Department of Physics, University of Maryland, \\College Park,
Maryland 20742, U.S.A.}

\end{center}

\vspace{1ex}

\begin{abstract}
It is possible to construct representations of the Lorentz group using
four-dimensional harmonic oscillators.  This allows us to construct
three-dimensional wave functions with the usual rotational symmetry
for space-like coordinates and one-dimensional wave function for
time-like coordinate.  It is then possible to construct a
representation of the Poincar\'e group for a massive particles having
the $O(3)$ internal space-time symmetry in its rest frame.  This
oscillator can also be separated into two transverse components and
the two-dimensional world of the longitudinal and time-like coordinates.
The transverse components remain unchanged under Lorentz boosts, while
it is possible to construct the squeeze representation of the $O(1,1)$
group in the space of the longitudinal and time-like coordinates.
While the squeeze representation forms the basic language for squeezed
states of light, it can be combined with the transverse components to
form the representation of the Poincar\`e group for relativistic
extended particles.
\end{abstract}

\section{Introduction}\label{intro}
Harmonic oscillators played the essential role in the development of
quantum mechanics.  From the mathematical point of view, the present
form of quantum mechanics did not advance too far from the oscillator
framework.  Thus, the first relativistic wave function has to be the
oscillator wave function.  With this point in mind, Dirac in 1945 wrote
down a normalizable four-dimensional wave function and attempted to
construct representations of the Lorentz group~\cite{dir45}, and
started a history of the oscillators which can be
Lorentz-boosted~\cite{yuka53,knp86}.

Let us consider the Minkowskian space consisting of the
three-dimensional space of $(x, y, z)$ and one time variable $t$.
Then the quantity $\left(x^{2} + y^{2} + z^{2} - t^{2}\right)$
remains invariant under Lorentz transformations.  On the other hand,
$\left(x^{2} + y^{2} + z^{2} + t^{2}\right)$ is not invariant.
Thus, the exponential form
\begin{equation}\label{ex1}
\exp\left(x^{2} + y^{2} + z^{2} - t^{2}\right)
\end{equation}
is a Lorentz-invariant quantity, while the form
\begin{equation}\label{ex2}
\exp\left(x^{2} + y^{2} + z^{2} + t^{2}\right)
\end{equation}
is not.  The Gaussian function of Eq.(\ref{ex2}) is localized in the
$t$ variable.  It was Dirac who wrote down this normalizable Gaussian
form in 1945~\cite{dir45}.  In 1963, the author of this report was
fortunate enough to hear directly from Prof. Dirac that the physics of
special relativity is much richer than writing down Lorentz-invariant
quantities.  This could mean that we should study more systematically
the above normalizable form under the influence of Lorentz boosts,
and this study has led to the observation that Lorentz boosts are
squeeze transformations~\cite{kn73,knp91}.

The exponential form of Eq.(\ref{ex1}) is Lorentz-invariant but cannot
be normalized.  This aspect was noted by Feynman {\it et al}. in their
1971 paper~\cite{fkr71}.  In their paper, Feynman {\it et al}. tell us
not to use Feynman diagrams but to use oscillator wave functions when
we approach relativistic bound states.  In this report, we shall use
much of the formalism given in Ref.~\cite{fkr71}, but not their
Lorentz-invariant wave functions which are not normalizable.

In this report, we discuss the oscillator representations of the
$O(3,1)$ and explain why this representation is adequate for internal
space-time symmetry of relativistic extended hadrons.  This group
has $O(1,1)$ as a subgroup which describes Lorentz boosts along a
given direction.  It is shown that Lorentz boosts are squeeze
transformations.  It is shown also that the infinite-dimensional
unitary representation of this squeeze group constitutes the
mathematical basis for squeezed states of light.

In Sec.~\ref{subg},  we discuss the three-parameter subgroups of the
six-parameter Lorentz group which leaves the four-momentum of the
particle invariant. These groups govern the internal space-time
symmetries of relativistic particles, and they are called Wigner's
little groups~\cite{wig39}.
In Sec.~\ref{covham}, it is shown that the covariant harmonic
oscillators constitute a representation of the Poincar\'e group for
relativistic extended particles.  If we add Lorentz boosts to
the oscillator formalism, the symmetry group becomes non-compact, and
its unitary representations become infinite-dimensional.
In Sec.~\ref{infinite}, we study an infinite-dimensional unitary
representation for the harmonic oscillator formalism.

\section{Subgroups of the Lorentz Group}\label{subg}
The Poincar\'e group is the group of inhomogeneous Lorentz
transformations, namely Lorentz transformations followed by space-time
translations.  This group is known as the fundamental space-time
symmetry group for relativistic particles.  This ten-parameter group
has many different representations.  For space-time symmetries, we study
first Wigner's little groups.  The little group is a maximal subgroup
of the six-parameter Lorentz group whose transformations leave the
four-momentum of a given particle invariant.  The little group therefore
governs the internal space-time symmetry of the particles.  Massive
particles in general are known as spin degrees of freedom.  Massless
particles in general have the helicity and gauge degrees of freedom.

Let $J_{i}$ be the three generators of the rotation group and $K_{i}$
be the three boost generators.  They then satisfy the commutation
relations
\begin{equation}
\left[J_{i}, J_{j}\right] = i \epsilon_{ijk} J_{k} ,  \qquad
\left[J_{i}, K_{j}\right] = i \epsilon_{ijk} K_{k} , \qquad
\left[K_{i}, K_{j}\right] = - i \epsilon_{ijk} J_{k} .
\end{equation}
The three-dimensional rotation group is a subgroup of the Lorentz group.
If a particle is at rest, we can rotate it without changing the
four-momentum.  Thus, the little group for massive particles is
the three-parameter rotation group.  If the particle is boosted, it
gains a momentum along the boosted direction.  If it is boosted along
the $z$ direction, the boost operator becomes
\begin{equation}
B_{3}(\eta) = \exp\left(-i\eta K_{3}\right) .
\end{equation}
The little group is then generated by
\begin{equation}
J'_{i} = B_{3}(\eta) J_{3} \left(B_{3}(\eta)\right)^{-1} .
\end{equation}
These boosted $O(3)$ generators satisfy the same set of commutation
relations as the set for $O(3)$.

\vspace{5mm}
\begin{quote}
Table I. Covariance of the energy-momentum relation, and covariance of
the internal space-time symmetry groups.  The quark model and the parton
model are two different manifestations of the same covariant entity.
\end{quote}

\begin{center}

\begin{tabular}{ccc}
Massive, Slow & COVARIANCE & Massless, Fast \\[2mm]\hline
{}&{}&{}\\
$E = p^{2}/2m$ & Einstein's $E = mc^{2}$ & $E = cp$ \\[4mm]\hline
{}&{}&{}  \\
$S_{3}$ & {}  &    $S_{3}$ \\ [-1mm]
{} & Wigner's Little Group & {} \\[-1mm]
$S_{1}, S_{2}$ & {} & Gauge Trans. \\[4mm]\hline
{}&{}&{}  \\
{}&{}&{}  \\ [-2mm]
Quarks & Covariant Harmonic Oscillators & Partons \\[8mm]\hline
\end{tabular}

\end{center}

\vspace{5mm}
If the parameter $\eta$ becomes infinite, the particle becomes like
a massless particle.  If we go through the contraction procedure
spelled out by Inonu and Wigner in 1953~\cite{inonu53}, the $O(3)$-like
little group becomes contracted to the $E(2)$-like little group for
massless particles generated by $J_{3}, N_{1}$, and
$N_{2}$~\cite{bacryc68,hks83pl},where
\begin{equation}
N_{1} = K_{1} - J_{2} , \qquad N_{2} = K_{2} + J_{1} .
\end{equation}
These $N$ generators are known to generate gauge transformations for
massless particles~\cite{janner71,hks82}.
Gauge transformations for spin-1 photons are
well known.  As for massless spin-1/2 particles, neutrino polarizations
are due to gauge invariance.

The transition from massive to massless particles is illustrated
in the second row of Table I.  In Sec. \ref{covham}, we shall
discuss how a massive particle with space-time extension be boosted
from its rest frame to an infinite-momentum frame. This aspect is
illustrated in the third row of Table I.

\section{Covariant Harmonic Oscillators}\label{covham}
If we construct a representation of the Lorentz group using normalizable
harmonic oscillator wave functions, the result is the covariant harmonic
oscillator formalism~\cite{knp86}.  The formalism constitutes a
representation of Wigner's O(3)-like little group for a massive particle
with internal space-time structure.  This oscillator formalism has been
shown to be effective in explaining the basic phenomenological features
of relativistic extended hadrons observed in high-energy laboratories.
In particular, the formalism shows that the quark model and Feynman's
parton picture~\cite{fey69} are two different manifestations of one
relativistic entity \cite{knp86,kim89}.

The essential feature of the covariant harmonic oscillator formalism is
that Lorentz boosts are squeeze transformations~\cite{kn73}.  In the
light-cone coordinate system, the boost transformation expands one
coordinate while contracting the other so as to preserve the product of
these two coordinate remains constant.  We shall show that the parton
picture emerges from this squeeze effect.

The covariant harmonic oscillator formalism has been discussed exhaustively
in the literature, and it is not necessary to give another full-fledged
treatment in the present paper.  We shall discuss here one of the most
puzzling problems in high-energy physics, namely whether quarks are partons.
It is now a well-accepted view that hadrons are bound states of quarks.
This view is correct if the hadron is at rest or nearly at rest.  On the
other hand, it appears as a collection of partons when it moves with a
speed very close to that of light.  This is called Feynman's parton picture
\cite{fey69}.

Let us consider a bound state of two particles.  For convenience, we shall
call the bound state the hadron, and call its constituents quarks.  Then
there is a Bohr-like radius measuring the space-like separation between the
quarks.  There is also a time-like separation between the quarks, and this
variable becomes mixed with the longitudinal spatial separation as the
hadron moves with a relativistic speed.  There are no quantum excitations
along the time-like direction.  On the other hand, there is the time-energy
uncertainty relation which allows quantum transitions.  It is possible to
accommodate these aspect within the framework of the present form of quantum
mechanics.  The uncertainty relation between the time and energy variables is
the c-number relation, which does not allow excitations along the time-like
coordinate.  We shall see that the covariant harmonic oscillator formalism
accommodates this narrow window in the present form of quantum mechanics.

Let us consider a hadron consisting of two quarks.  If the space-time
position of two quarks are specified by $x_{a}$ and $x_{b}$ respectively,
the system can be described by the variables
\begin{equation}
X = (x_{a} + x_{b})/2 , \qquad x = (x_{a} - x_{b})/2\sqrt{2} .
\end{equation}
The four-vector $X$ specifies where the hadron is located in space and time,
while the variable $x$ measures the space-time separation between the
quarks.  In the convention of Feynman {\it et al.}~\cite{fkr71}, the internal
motion of the quarks bound by a harmonic oscillator potential of unit
strength can be described by the Lorentz-invariant equation
\begin{equation}
{1\over 2}\left\{x^{2}_{\mu} - {\partial ^{2} \over \partial x_{\mu}^{2}}
\right\} \psi(x)= \lambda \psi(x) .
\end{equation}
We use here the space-favored metric: $x^{\mu} = (x, y, z, t)$.

It is possible to construct a representation of the Poincar\'e group
from the solutions of the above differential equation~\cite{knp86}.
If the hadron is at rest, the solution should take the form
\begin{equation}
\psi(x,y,z,t) = \phi(x,y,z) \left({1\over \pi}\right)^{1/4}
\exp \left(-t^{2}/2 \right) ,
\end{equation}
where $\phi(x,y,z)$ is the wave function for the three-dimensional
oscillator.  If we use the spherical coordinate system, this wave
function will carry appropriate angular momentum quantum numbers.
Indeed, the above wave function constitutes a representation of Wigner's
$O(3)$-like little group for a massive particle~\cite{knp86}. There are
no time-like excitations, and this is consistent with our observation
of the real world.  It was Dirac who noted first this space-time
asymmetry in quantum mechanics~\cite{dir27}.  However, this asymmetry
is quite consistent with the $O(3)$ symmetry of the little group for
hadrons.

Since the three-dimensional oscillator differential equation is
separable in both the spherical and Cartesian coordinate systems, the
spherical form of $\phi(x,y,z)$ consists of Hermite polynomials of
$x, y$, and $z$.  If the Lorentz boost is made along the $z$ direction,
the $x$ and $y$ coordinates are not affected, and can be dropped from
the wave function.  The wave function of interest can be written as
\begin{equation}
\psi^{n}(z,t) = \left({1\over \pi}\right)^{1/4}\exp \pmatrix{-t^{2}/2}
\phi_{n}(z) ,
\end{equation}
with
\begin{equation}\label{1dwf}
\phi_{n}(z) = \left({1 \over \pi n!2^{n}} \right)^{1/2} H_{n}(z)
\exp (-z^{2}/2) ,
\end{equation}
where $\psi^{n}(z)$ is for the $n$-th excited oscillator state.
The full wave function $\psi^{n}(z,t)$ is
\begin{equation}\label{wf}
\psi^{n}_{0}(z,t) = \left({1\over \pi n! 2^{n}}\right)^{1/2} H_{n}(z)
\exp \left\{-{1\over 2}\left(z^{2} + t^{2} \right) \right\} .
\end{equation}
The subscript 0 means that the wave function is for the hadron at rest.  The
above expression is not Lorentz-invariant, and its localization undergoes a
Lorentz squeeze as the hadron moves along the $z$ direction~\cite{knp86}.

It is convenient to use the light-cone variables to describe Lorentz boosts.
The light-cone coordinate variables are
\begin{equation}
u = (z + t)/\sqrt{2} , \qquad v = (z - t)/\sqrt{2} .
\end{equation}
In terms of these variables, the Lorentz boost along the $z$
direction,
\begin{equation}
\pmatrix{z' \cr t'} = \pmatrix{\cosh \eta & \sinh \eta \cr \sinh \eta &
\cosh \eta}\pmatrix{z \cr t} ,
\end{equation}
takes the simple form
\begin{equation}\label{lorensq}
u' = e^{\eta} u , \qquad v' = e^{-\eta} v ,
\end{equation}
where $\eta $ is the boost parameter and is $\tanh ^{-1}(v/c)$.

The wave function of Eq.(\ref{wf}) can be written as
\begin{equation}\label{10}
\psi^{n}_{o}(z,t) = \psi^{n}_{0}(z,t)
= \left({1 \over \pi n!2^{n}} \right)^{1/2} H_{n}\left((u + v)/\sqrt{2}
\right) \exp \left\{-{1\over 2} (u^{2} + v^{2}) \right\} .
\end{equation}
If the system is boosted, the wave function becomes
\begin{equation}\label{11}
\psi^{n}_{\eta}(z,t) = \left({1 \over \pi n!2^{n}} \right)^{1/2}
H_{n} \left((e^{-\eta}u + e^{\eta}v)/\sqrt{2} \right)
\times \exp \left\{-{1\over 2}\left(e^{-2\eta}u^{2} +
e^{2\eta}v^{2}\right)\right\} .
\end{equation}

In both Eqs. (\ref{10}) and (\ref{11}), the localization property of the wave
function in the $u v$ plane is determined by the Gaussian factor, and it
is sufficient to study the ground state only for the essential feature of
the boundary condition.  The wave functions in Eq.(\ref{10}) and
Eq.(\ref{11}) then respectively become
\begin{equation}\label{13}
\psi^{0}(z,t) = \left({1 \over \pi} \right)^{1/2}
\exp \left\{-{1\over 2} (u^{2} + v^{2}) \right\} .
\end{equation}
If the system is boosted, the wave function becomes
\begin{equation}\label{14}
\psi_{\eta}(z,t) = \left({1 \over \pi}\right)^{1/2}
\exp \left\{-{1\over 2}\left(e^{-2\eta}u^{2} +
e^{2\eta}v^{2}\right)\right\} .
\end{equation}
We note here that the transition from Eq.(\ref{13}) to Eq.(\ref{14}) is a
squeeze transformation.  The wave function of Eq.(\ref{13}) is distributed
within a circular region in the $u v$ plane, and thus in the $z t$ plane.
On the other hand, the wave function of Eq.(\ref{14}) is distributed in an
elliptic region.

\section{Unitary Infinite-dimensional Representation}\label{infinite}
Let us go back to Eq.(\ref{10}) and Eq.(\ref{11}).  We are now
interested in writing them in terms of the one-dimensional oscillator
wave functions given in Eq.(\ref{1dwf}).  After some standard
calculations~\cite{knp86}, we can write the squeezed wave function as
\begin{equation}\label{power}
\psi^{n}_{\eta}(z,t) = \left({1\over\cosh\eta}\right)^{n+1}\sum^{}_{k}
\left((n + k)!\over n!k!\right)^{1/2}(\tanh\eta)^{k} \phi_{n+k}(z)
\phi_{n}(t) .
\end{equation}
If the parameter $\eta$ becomes zero, this form becomes the rest-frame
wave function of Eq.(\ref{10}).

It is sometimes more convenient to use the parameter $\beta$ defined
as
\begin{equation}
\beta = \tanh\eta .
\end{equation}
This parameter is the speed of the hadron divided by the speed of light.
In terms of this parameter, the expression of Eq.(\ref{power}) can be
written as
\begin{equation}\label{power2}
\psi^{n}_{\eta}(z,t) = \left(1 - \beta^{2}\right)^{(n+1)/2}\sum^{}_{k}
\left((n + k)!\over n!k!\right)^{1/2} \beta^{k} \phi_{n+k}(x)
\phi_{n}(t) .
\end{equation}
If we take the integral
\begin{equation}\label{integ}
\int |\psi^{n}_{\eta}(z,t)|^{2} dz dt
= \left(1 - \beta^{2}\right)^{n+1} \sum^{}_{k}
\left((n + k)!\over n!k!\right) \beta^{2k} .
\end{equation}
The sum in the above expression is the same as the binomial expansion
of
$$
\left(1 - \beta^{2}\right)^{-(n+1)}.
$$
Thus, the right hand side of Eq.(\ref{integ}) is 1.  The power series
expansion of Eq.(\ref{power}) reflects a well-known but hard-to-prove
mathematical theorem that unitary representations of non-compact groups
are infinite-dimensional.  The Lorentz group is a non-compact group.

If $n = 0$, the above form becomes simplified to
\begin{equation}\label{power0}
\psi_{\eta}(z,t) = \left(1 - \beta^{2}\right)^{1/2}\sum^{}_{k}
\beta^{k} \phi_{k}(z) \phi_{k}(t) .
\end{equation}
This is the power series expansion of Eq.(\ref{14}).  This relatively
simple form is very useful in many other branches of physics.

It is well known that the mathematics of the Fock space in quantum
field theory is that of harmonic oscillators.  Among them, the
coherent-state representation occupies a prominent place because
it is the basic language for laser optics.  Recently, two photon
coherent states have been observed and the photon distribution is
exactly like that of the wave function given in Eq.(\ref{power}).
These coherent states are commonly called squeezed states.  It is
very difficult to see why the word ``squeeze'' has to be associated
with the power series expansion given in Eq.(\ref{power}) or
Eq.(\ref{power2}).  It is however quite clear from the expression of
Eq.(\ref{14}) that Gaussian distribution is squeezed.  Thus, the
above representation tells us how squeezed states are squeezed.

Next, let us briefly discuss the role of this infinite dimensional
representation in understanding the density matrix.  For simplicity,
we shall work with the squeezed ground-state wave function.  From the
wave function of Eq.(\ref{power0}), we can construct the pure-state
density matrix
\begin{equation}\label{pure}
\rho_{\eta}(z,t;z',t') = \psi_{\eta}(z,t)\psi_{\eta}(z',t') ,
\end{equation}
which satisfies the condition $\rho^{2} = \rho $:
\begin{equation}
\rho_{\eta}(z,t;z',t') = \int \rho_{\eta}(z,t;z'',t'')
\rho_{\eta}(z'',t'';z',t') dz'' dt'' .
\end{equation}
However, there are at present no measurement theories which accommodate
the time-separation variable t.  Thus, we can take the trace of the
$\rho$ matrix with respect to the $t$ variable.  Then the resulting
density matrix is
\begin{equation}\label{densi}
\rho_{\eta}(z,z') = \int \psi_{\eta}(z,t)
\left\{\psi_{\eta}(z',t)\right\}^{*} dt
= \left(1 - \beta^{2}\right)\sum^{}_{k}
\beta^{2k}\phi_{k}(z)\phi^{*}_{k}(z') .
\end{equation}
It is of course possible to compute the above integral using the
analytical expression given in Eq.(\ref{14}).  The result is
\begin{equation}
\rho_{\eta}(z,z') = \left({1\over \pi \cosh 2\eta} \right)^{1/2}
\exp\left\{-{1\over 4}[(z + z')^{2}/\cosh 2\eta
+ (z - z')^{2}\cosh 2\eta] \right\} .
\end{equation}
This form of the density matrix satisfies the trace condition
\begin{equation}
\int \rho(z,z) dz = 1 .
\end{equation}
The trace of this density matrix is one, but the trace of $\rho^{2}$ is
less than one, as
\begin{equation}
Tr\left(\rho^{2}\right) = \int \rho^{n}_{\eta}(z,z')
\rho^{n}_{\eta}(z',z) dz'dz
= \left(1 - \beta^{2}\right)^{2} \sum^{}_{k} \beta^{4k} ,
\end{equation}
which is less than one and is $1/\left(1 + \beta^{2}\right)$.  This is
due to the fact that we do not know how to deal with the time-like
separation in the present formulation of quantum mechanics.  Our
knowledge is less than complete.

The standard way to measure this ignorance is to calculate the
entropy defined as~\cite{neu32,wiya63}
\begin{equation}
S = - Tr\left(\rho\ln(\rho)\right) .
\end{equation}
If we pretend to know the distribution along the time-like
direction and use the pure-state density matrix given in Eq.(\ref{pure}),
then the entropy is zero.  However, if we do not know how to deal with
the distribution
along $t$, then we should use the density matrix of Eq.(\ref{densi}) to
calculate the entropy, and the result is~\cite{kiwi90pl}
\begin{equation}
S = 2 [(\cosh\eta)^{2}\ln (\cosh\eta) - (\sinh\eta)^{2}\ln(\sinh\eta)] .
\end{equation}
In terms of the velocity parameter, this expression can be written as
\begin{equation}
S = {1\over 1 - \beta^{2}}\ln{1\over 1 - \beta^{2}} -
{\beta \over 1 - \beta^{2}}\ln{\beta \over 1 - \beta^{2}} .
\end{equation}
From this we can derive the hadronic temperature~\cite{hkn90pl}.

In this report, the time-separation variable $t$ played the role of
an unmeasurable variable.  The use of an unmeasurable variable as
a ``shadow'' coordinate is not new in physics and is of current
interest~\cite{ume82}.  Feynman's book on statistical mechanics
contains the following paragraph~\cite{fey72}.

{\it When we solve a quantum-mechanical problem, what we really do
is divide the universe into two parts - the system in which we are
interested and the rest of the universe.  We then usually act as if
the system in which we are interested comprised the entire universe.
To motivate the use of density matrices, let us see what happens
when we include the part of the universe outside the system.}

In the present paper, we have identified Feynman's rest of the
universe as the time-separation coordinate in a relativistic two-body
problem.  Our ignorance about this coordinate leads to a density matrix
for a non-pure state, and consequently to an increase of entropy.

\end{document}